\documentclass[prl, twocolumn, amsmath, amssymb, showpacs]{revtex4-1}
\usepackage{times}
\usepackage{verbatim}
\usepackage{graphicx}
\usepackage{graphics}
\usepackage{amsmath}
\usepackage{amssymb}
\begin{document}

\title{Effect of Cleaving Temperature on the Surface and Bulk Fermi Surface of Sr$_2$RuO$_4$ Investigated by High Resolution Angle-Resolved Photoemission}

\author{Shanyu Liu$^{1}$, Wentao Zhang$^{1}$, Hongming Weng$^{2}$, Lin Zhao$^{1}$, Haiyun Liu$^{1}$, Xiaowen Jia$^{1}$, Guodong Liu$^{1}$, Xiaoli Dong$^{1}$, Jun Zhang$^{1}$, Z. Q. Mao$^{3}$, Chuangtian Chen$^{4}$, Zuyan Xu$^{4}$,  Xi Dai$^{2}$, Zhong Fang$^{2}$ and X. J. Zhou$^{1}$$^{*}$}

\affiliation{
\\$^1$National Laboratory for Superconductivity, Beijing National Laboratory for Condensed Matter Physics, Institute of Physics, Chinese Academy of Sciences, Beijing 100190, China
\\$^2$Beijing National Laboratory for Condensed Matter Physics, Institute of Physics, Chinese Academy of Sciences, Beijing 100190, China
\\$^3$Department of Physics and Engineering Physics, Tulane University, New Orleans, Louisianan 70118, USA
\\$^4$Technical Institute of Physics and Chemistry, Chinese Academy of Sciences, Beijing 100190, China}

\begin{abstract}
High resolution angle-resolved photoemission measurements are carried out to systematically investigate the effect of cleaving temperature on the electronic structure and Fermi surface of Sr$_2$RuO$_4$. Different from previous reports that high  cleaving temperature can suppress surface Fermi surface,  we find that the surface Fermi surface remains obvious and strong in Sr$_2$RuO$_4$ cleaved at high temperature, even at room temperature. This indicates that cleaving temperature is not a key effective factor in suppressing the surface bands. On the other hand, in the aged surface of Sr$_2$RuO$_4$ that is cleaved and held for a long time, the bulk bands can be enhanced. We have also carried out laser ARPES measurements on Sr$_2$RuO$_4$ by using vacuum ultra-violet laser (photon energy at 6.994 eV) and found an obvious enhancement of bulk bands even for samples cleaved at low temperature. These information are important in realizing an effective approach in manipulating and detecting the surface and bulk electronic structure of Sr$_2$RuO$_4$. In particular, the enhancement of bulk sensitivity, together with its super-high instrumental resolution of VUV laser ARPES, will be advantageous in investigating fine electronic structure and superconducting properties of Sr$_2$RuO$_4$ in the future.
\end{abstract}
\maketitle


Sr$_2$RuO$_4$ is an unconventional superconductor that has attracted much attention\cite{SROReview,YMaenoReview,011009} since the discovery of superconductivity in 1994\cite{MaenoNature} with a transition temperature T$_c$ at 1.5 K\cite{sample}. So far, this is the only superconductor without copper that has a layered perovskite crystal structure identical to that of copper-oxide (La,Sr)$_2$CuO$_4$ high temperature  superconductor\cite{1}. Theoretical expectations\cite{TMRice,CHonerkamp} and experimental measurements \cite{Luke, KIshida, KDNelson, JXia,Kidwingira} indicate that Sr$_2$RuO$_4$ is an unconventional superconductor with a spin-triplet pairing and a possible p$_x$+ip$_y$ superconducting order parameter.  Its normal state above T$_c$ exhibits Fermi-liquid like behaviors\cite{BergemannReview}.  Most recently, Sr$_2$RuO$_4$ is classified as a topological superconductor which is closely related with Majorana Fermions and non-Abelian statistics\cite{SCZhangReview,DaSarma}.

The electronic structure and physical properties of Sr$_2$RuO$_4$ are dictated by the Ru 4d orbitals\cite{LDATOguchi, LDADJSingh, IIMazin1, IIMazin2}. The low-lying electronic states originate from the hybridization between the Ru 4$\it{d}$$_{xy}$, 4$\it{d}$$_{xz}$, 4$\it{d}$$_{yz}$ orbitals and the oxygen 2$\it{p}$ orbitals. The quasi-two-dimensional nature of the crystal structure gives rise to nearly two-dimensional Fermi surface sheets. Among them , the $\it{d}$$_{xy}$ orbital produces an almost cylindrical Fermi surface sheet around the center of the Brillouin zone (usually denoted as $\gamma$ sheet, Fig. 1b). On the other hand, the nearly one-dimensional nature of the $\it{d}$$_{xz}$ and $\it{d}$$_{yz}$  orbitals produces nearly planar Fermi surface sheets which run perpendicular to x-axis and y-axis, respectively. The hybridization between the  $\it{d}$$_{xz}$ and $\it{d}$$_{yz}$ orbital-generated bands gives rise to a hole-like Fermi surface sheet enclosed around M($\pi$,$\pi$) point (denoted as $\alpha$ sheet) and an electron-like Fermi surface sheet around the $\Gamma$(0,0) (denoted as $\beta$ sheet, Fig. 1b). Such an electronic structure of Sr$_2$RuO$_4$ has been well described by band structure calculations\cite{LDATOguchi, LDADJSingh, IIMazin1, IIMazin2} and experimentally confirmed by the quantum oscillation measurements\cite{Mackenzie,Bergemann}.

Angle-resolved photoemission spectroscopy (ARPES) is a powerful tool to directly determine the electronic structure and Fermi surface of materials\cite{DamascelliReview}. However, the ARPES measurements on Sr$_2$RuO$_4$ experienced a complicated and controversial process. Earlier ARPES results on Sr$_2$RuO$_4$\cite{Lu,Yokoya,AVPuchkov} were not compatible with the band structure calculations and quantum oscillation measurements, rendering people doubt whether ARPES, as a surface sensitive technique, is suitable for studying bulk electronic properties\cite{MackenzieComment}. It turned out that  Sr$_2$RuO$_4$ exhibits a ${\sqrt2}$$\times$${\sqrt2}$ surface reconstruction arising from the rotation of the RuO$_6$ octahedra in the surface layer\cite{Matzdorf}. This surface reconstruction made it possible to resolve the controversy surrounding the electronic structure of Sr$_2$RuO$_4$\cite{DamascelliPRL, KShenSurface}. The occurrence of surface reconstruction produces two effects. First, in addition to the original one set of bulk Fermi surface sheets ($\alpha_B$, $\beta_B$ and $\gamma_B$), it produces an additional set of surface Fermi surface sheets ($\alpha_S$, $\beta_S$ and $\gamma_S$). Second, the ${\sqrt2}$$\times$${\sqrt2}$ surface reconstruction produces two additional sets of Fermi surface sheets: surface shadow bands (S$\alpha_S$, S$\beta_S$ and S$\gamma_S$) formed by inverting the original surface Fermi surface sheets ($\alpha_S$, $\beta_S$ and $\gamma_S$) with respect to the ($\pi$,0)$-$(0,$\pi$) line, and bulk shadow bands (S$\alpha_B$, S$\beta_B$ and S$\gamma_B$) formed also by inverting the original bulk Fermi surface sheets ($\alpha_B$, $\beta_B$ and $\gamma_B$) with respect to the ($\pi$,0)$-$(0,$\pi$) line.  Considering the four sets of Fermi surface sheets shown in Fig. 1d, the bulk Fermi surface picked up from the ARPES measurements\cite{DamascelliPRL} becomes consistent with the band structure calculations and quantum oscillation measurements.

\begin{figure}[t]
\begin{center}
\includegraphics[width=0.92\columnwidth,angle=0]{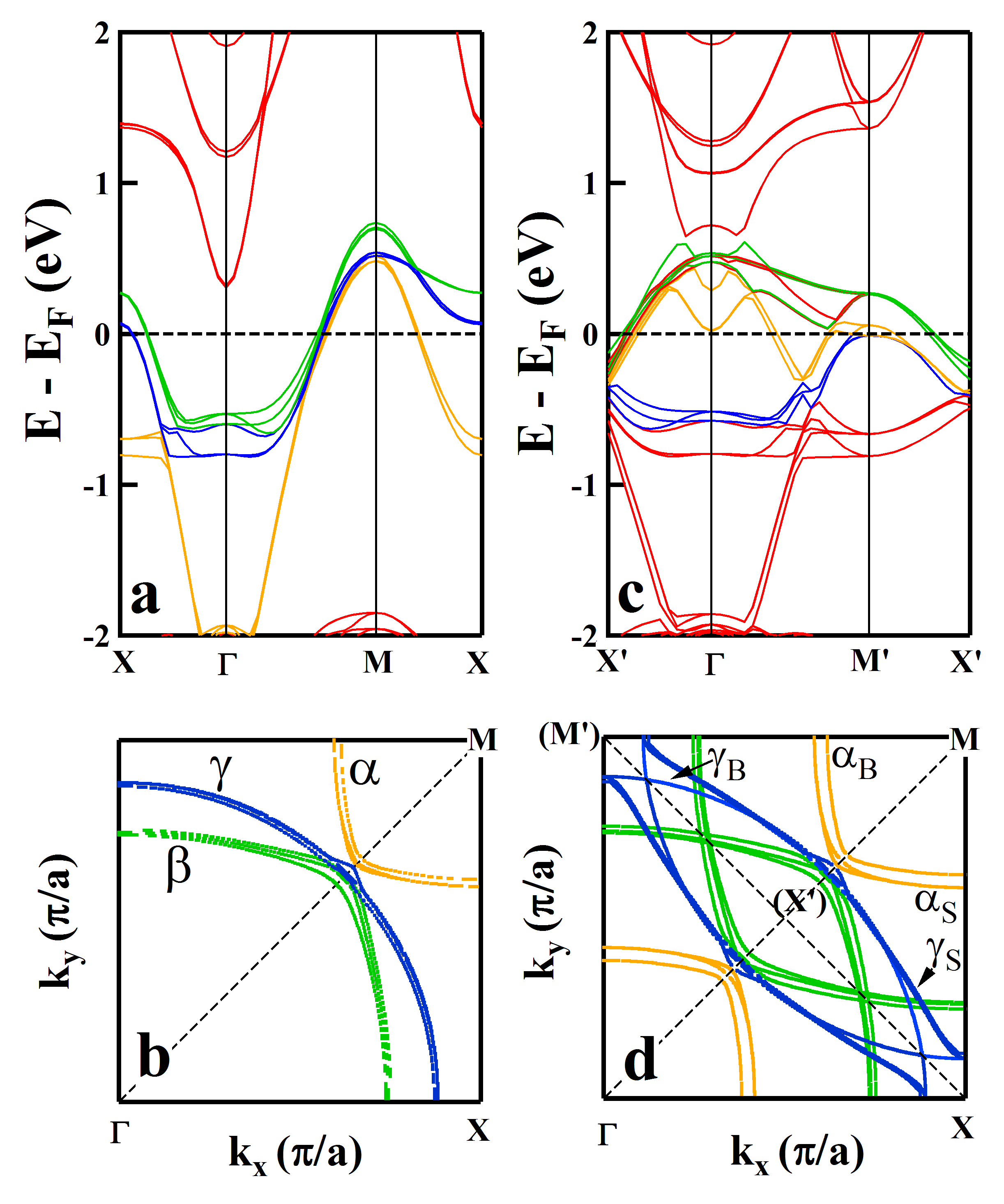}
\end{center}
\caption{LDA calculations of band structure and Fermi surface of Sr$_2$RuO$_4$. Repeated three-slab of Sr$_2$RuO$_4$ is used with the middle slab representing the bulk electronic structure while the top and bottom ones representing surface.  (a). Calculated band structure without considering any surface constructions.  (b). Fermi surface corresponding to band structure of (a) in one quadrant.  (c). Band structure with the top and bottom surface Ru-O octahedra rotating by 6 degrees. (d)  Fermi surface corresponding to band structure of (c). Note that here because of the ${\sqrt2}$$\times$${\sqrt2}$ surface reconstruction, the real Brillouin zone is reduced by half.}
\end{figure}

The understanding of the surface crystal structure and surface electronic structure of Sr$_2$RuO$_4$ is crucial for both fundamental studies of Sr$_2$RuO$_4$ and its potential applications as a topological material in quantum computing\cite{SCZhangReview,DaSarma}. First, as mentioned above, the uncovering of surface reconstruction\cite{Matzdorf} is crucial in understanding the ARPES-measured Fermi surface of Sr$_2$RuO$_4$\cite{DamascelliPRL};  Second, it remains unclear whether Sr$_2$RuO$_4$ can support surface magnetism, as expected from theory\cite{Matzdorf} and examined by ARPES measrement\cite{KShenSurface}. Third, the detection of possible edge state\cite{Kashiwaya} associated with Sr$_2$RuO$_4$ as a topological superconductor asks for a comprehensive understanding and control of its surface properties. In many cases to investigate the canonical Fermi liquid behavior or many-body effects in Sr$_2$RuO$_4$, it is preferable to suppress the surface state to make the bulk state dominant\cite{DamascelliPRL,Iwashawa,NIngle,TKidd}. One way used is to cleave the Sr$_2$RuO$_4$ sample at high temperature (such as 180 K as in \cite{DamascelliPRL}) which was found to cause suppression of the surface Fermi surface sheets. However, whether the cleaving temperature is a key controlling factor and the mechanism of such a surface band suppression remain unclear\cite{Pennec}.




In this paper, we report systematic investigation on the influence of cleaving temperature on the electronic structure and Fermi surface of Sr$_2$RuO$_4$.  We find that the surface Fermi surface remains obvious and strong in samples cleaved at high temperature, even at room temperature. This is different from the previous claim\cite{DamascelliPRL}. Our results indicate that cleaving temperature is not a key effective factor to suppress the surface bands. On the other hand, in the aged surface of Sr$_2$RuO$_4$ cleaved and held for a long time, the bulk bands can be strongly enhanced. We also carried out laser ARPES measurements on Sr$_2$RuO$_4$ by using vacuum ultra-violet laser (photon energy at 6.994 eV)\cite{Liu2008} and found an obvious enhancement of bulk bands even for samples cleaved at low temperature. These results are important in finding out an effective way to manipulate and detect the surface and bulk electronic structure of Sr$_2$RuO$_4$.

Angle-resolved photoemission measurements on Sr$_2$RuO$_4$ were carried out on our lab ARPES system equipped with a R4000 electron energy analyzer together with both a helium discharge lamp ($\it{h}\nu$=21.218eV) and a vacuum ultraviolet (VUV)  laser ($\it{h}\nu$=6.994 eV)  as light sources\cite{Liu2008}. The energy resolution was set at 10 meV and 2 meV for helium lamp ARPES and VUV laser ARPES measurements, respectively. The angular resolution is $\sim$0.3$^{\circ}$. The Fermi level is determined by referencing the Fermi edge of a clean polycrystalline gold which is electronically connected to the sample. The Sr$_2$RuO$_4$ sample was grown by traveling solvent floating zone method and has a superconducting transition at T$_c$ = 1.5 K with a sharp transition width of $\sim$0.1 K. All samples were cleaved {\it in situ} and measured in vacuum with a base pressure better than 5$\times$10$^{-11}$ Torr.

\begin{figure}[t]
\begin{center}
\includegraphics[width=0.94\columnwidth,angle=0]{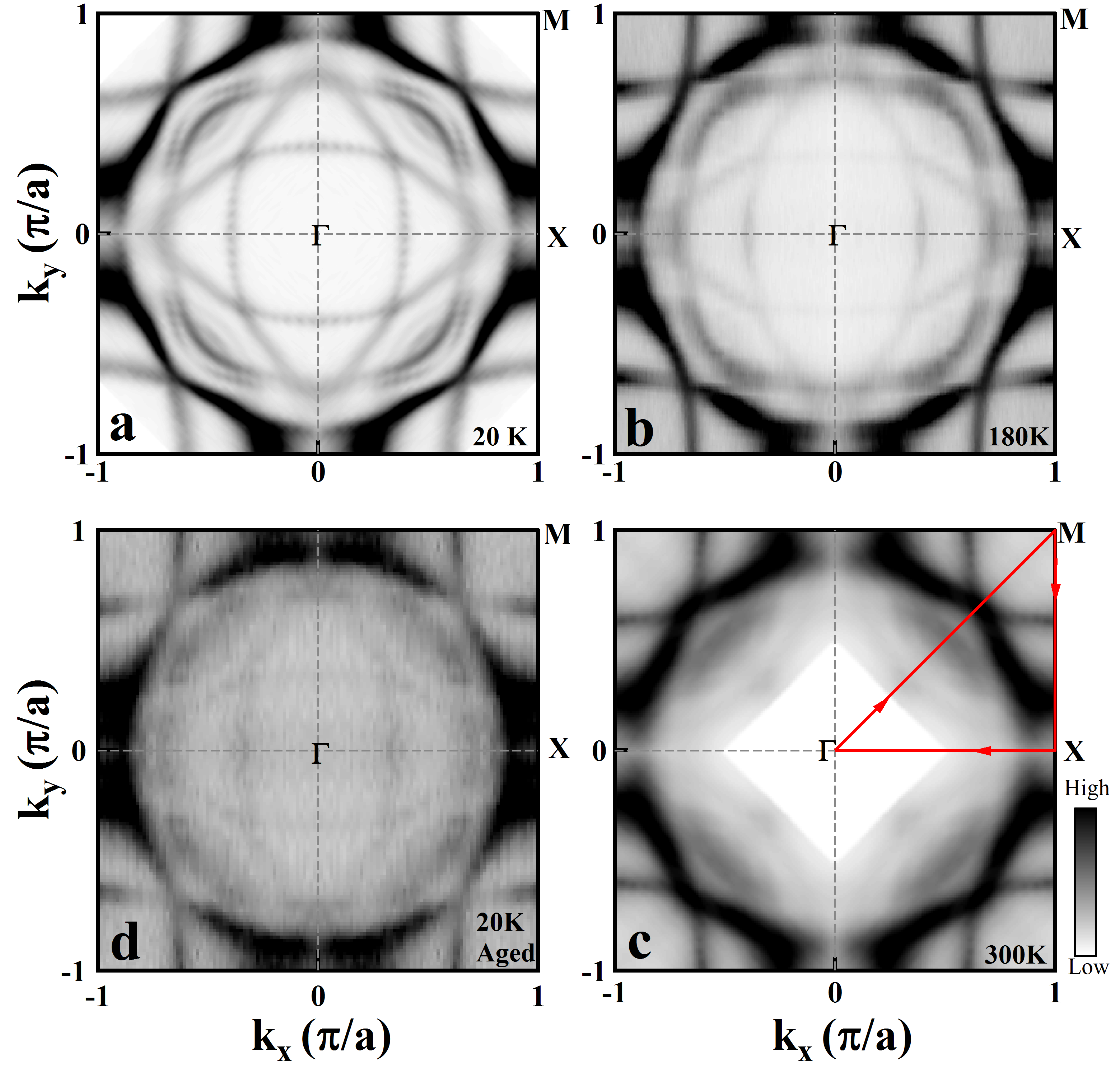}
\end{center}
\caption{ARPES measured Fermi surface of Sr$_2$RuO$_4$ at 20 K. (a). The sample was cleaved at 20 K and then immediately measured at 20 K;  (b). The sample was first cleaved at 180 K and then cooled down immediately to low temperature and measured at 20 K;  (c). The sample was first cleaved at 300 K and then measured immediately at 20 K;  (d). The sample was first cleaved at 20 K and measured after nearly 17 hours at 20 K. During this period, the sample was slowly warmed up to 100 K and then cooled down. The red lines in (c) shows the location of momentum cuts for the bands in Fig. 3.
}
\end{figure}

\begin{figure}[b]
\begin{center}
\includegraphics[width=0.95\columnwidth,angle=0]{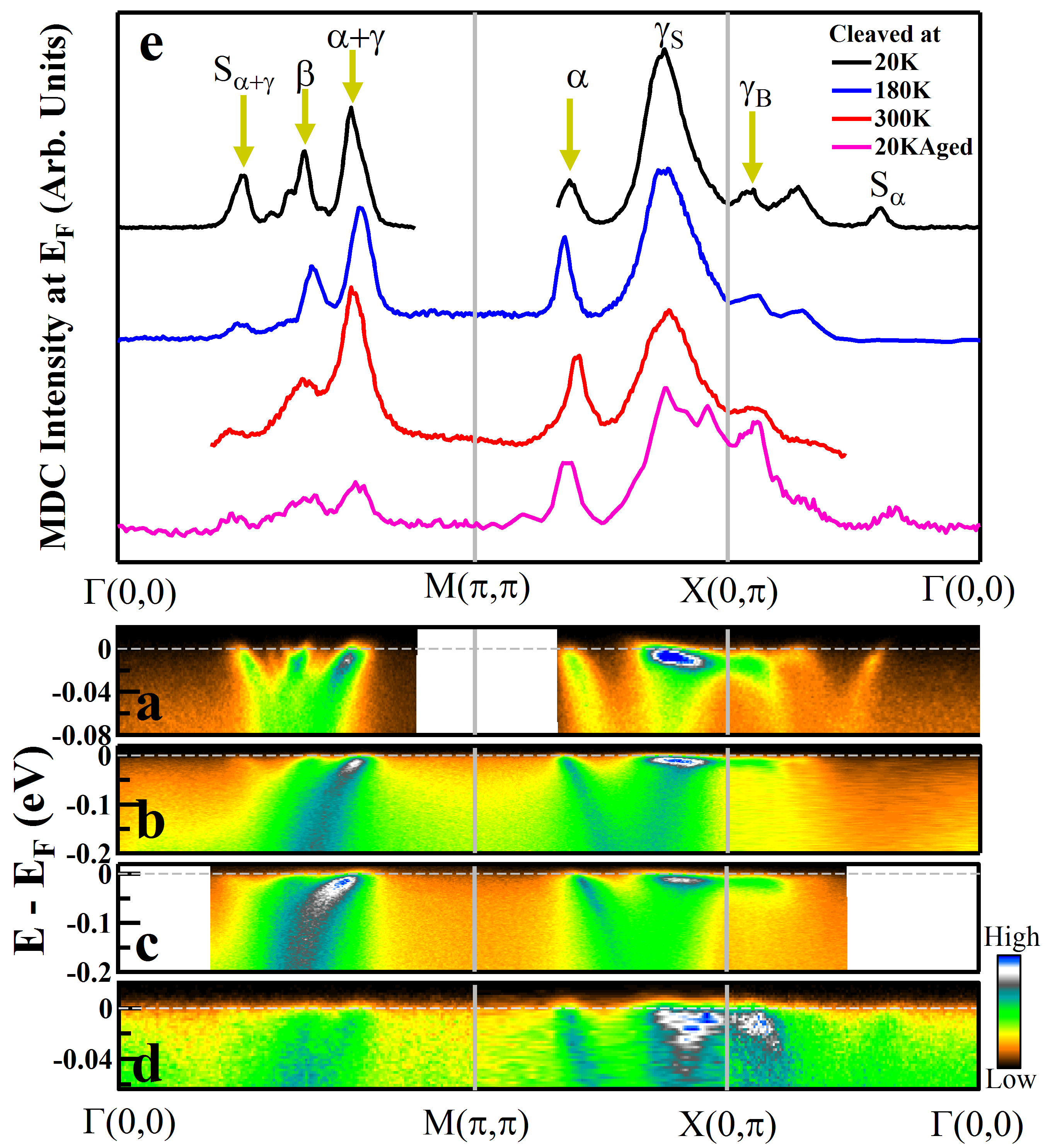}
\end{center}
\caption{Band structure and MDCs of Sr$_2$RuO$_4$ along $\Gamma$(0,0)-M($\pi$,$\pi$)-X(0,$\pi$)-$\Gamma$(0,0) high symmetry cuts, as shown by red lines in Fig. 2c. (a), (b), (c) and (d) show band structure for the Sr$_2$RuO$_4$ samples cleaved at 20 K, 180 K, 300 K and at 20 K but aged, respectively.  The corresponding momentum distribution curves (MDCs) are shown in (e). All the samples were measured at 20 K.
}
\end{figure}

Figure 2 shows ARPES-measured Fermi surface of Sr$_2$RuO$_4$ samples treated with different conditions; the corresponding energy bands along several high symmetry lines are shown in Fig. 3. There are two signatures to qualitatively compare the contributions of bulk and surface components.  One is the relative intensities of the bulk $\gamma_B$ band and surface $\gamma_S$ band. As seen in Fig. 1d, the bulk $\gamma_B$ band intersects with the $\Gamma$(0,0)$-$X(0,$\pi$) line while the surface $\gamma_S$ band intersects with the X(0,$\pi$)$-$M($\pi$,$\pi$) line; their relative intensity ratio thus provides a measure on the contributions of the bulk and surface bands. Another signature is the relative intensities of the shadow bands and their original bands. Because the shadow bands from the surface Fermi surface sheets are expected to be much stronger than those from the bulk Fermi surface sheets, their presence and relative intensity can give another measure on the contributions of the bulk and surface bands. It is clear that the surface components are dominant in the measured Fermi surface of the Sr$_2$RuO$_4$ sample freshly cleaved at low temperature (20 K, Fig. 2a and Fig. 3a), judging from both the relative intensity between the $\gamma_B$ band and the $\gamma_S$ band (Fig. 3e), and the relative intensity between the S$_{\alpha+\gamma}$ band(s) and the $\gamma_B$ band (Fig. 3e).  The surface components remain dominant for the Sr$_2$RuO$_4$ samples cleaved at high temperatures (180 K and 300 K, Figs. 2 and 3). The surface $\gamma_S$ remains strong and the bulk $\gamma_B$ remains weak for the samples cleaved at high temperatures (Fig. 3e). In addition, the existence of S$_{\alpha+\gamma}$ is clear. These indicate that cleaving Sr$_2$RuO$_4$ at high temperatures, even at room temperature, does not result in significant suppression of the surface bands, contrary to the previous report\cite{DamascelliPRL}. Therefore, the cleaving temperature is not a key factor to effectively suppress the surface bands; its effect in the previous work\cite{DamascelliPRL} might be due to other factors such as a particular gas component in the measurement chamber. For samples cleaved at high temperatures, the overall signal gets weak. Moreover, the measured Fermi surface sheets become blurred with some fine features clearly observed for low temperature cleaved sample (Fig. 2a) not resolvable for high temperature cleaved ones (Figs. 2b and 2c). These are additional disadvantages caused by high temperature cleaving.

Sample ageing, i.e., holding the freshly cleaved sample for a long period of time, can help suppress the surface bands. As shown in Fig. 2d and Figs. 3d and 3e, for the Sr$_2$RuO$_4$ sample cleaved at 20 K and held in ultra-high vacuum for 17 hours, the bulk bands are obviously enhanced. This can be judged by comparing the relative intensity of the bulk $\gamma_B$ band and surface $\gamma_S$ band in the MDC curve (Fig. 3e) for the aged sample which gets larger than that of a freshly cleaved sample. Further increase of the surface holding time may further enhance the bulk bands although we note that the overall signal is also significantly reduced in the aged samples.

It is known that ARPES is a surface sensitive technique.  For the helium discharge lamp with a photon energy at 21.2 eV, the estimated photoelectron escape depth is on the order of 5$\sim$10 ${\AA}$ that is just the top one or two layers near surface\cite{Seah1979}. As the photon energy decreases, the photoelectron escape depth is expected to increase to nearly 30 ${\AA}$ for the VUV laser with a photon energy of 6.994 eV\cite{Seah1979,Liu2008}. This prompts us to investigate the electronic structure of Sr$_2$RuO$_4$ by utilizing our VUV laser ARPES.  Fig.4 shows the Fermi surface measured with the VUV laser, and the band structure along several momentum cuts. It demonstrates that laser ARPES is a powerful tool to study Sr$_2$RuO$_4$. First, our VUV laser ARPES has the capability to reach the X($\pi$,0) point.  As shown in Fig. 4, in the measured momentum range, we clearly resolve $\alpha$ Fermi surface sheet and $\gamma$ Fermi surface sheet. Second, the maximum intensity of the bulk $\gamma_B$ sheet is comparable to that of the surface $\gamma_B$ sheet (Figs. 4c3 and 4c4). Compared with Fig. 3, it is clear that, even for the sample cleaved at 20 K, the bulk bands are strongly enhanced in our VUV laser-based ARPES measurements.

\begin{figure}[t]
\begin{center}
\includegraphics[width=0.94\columnwidth,angle=0]{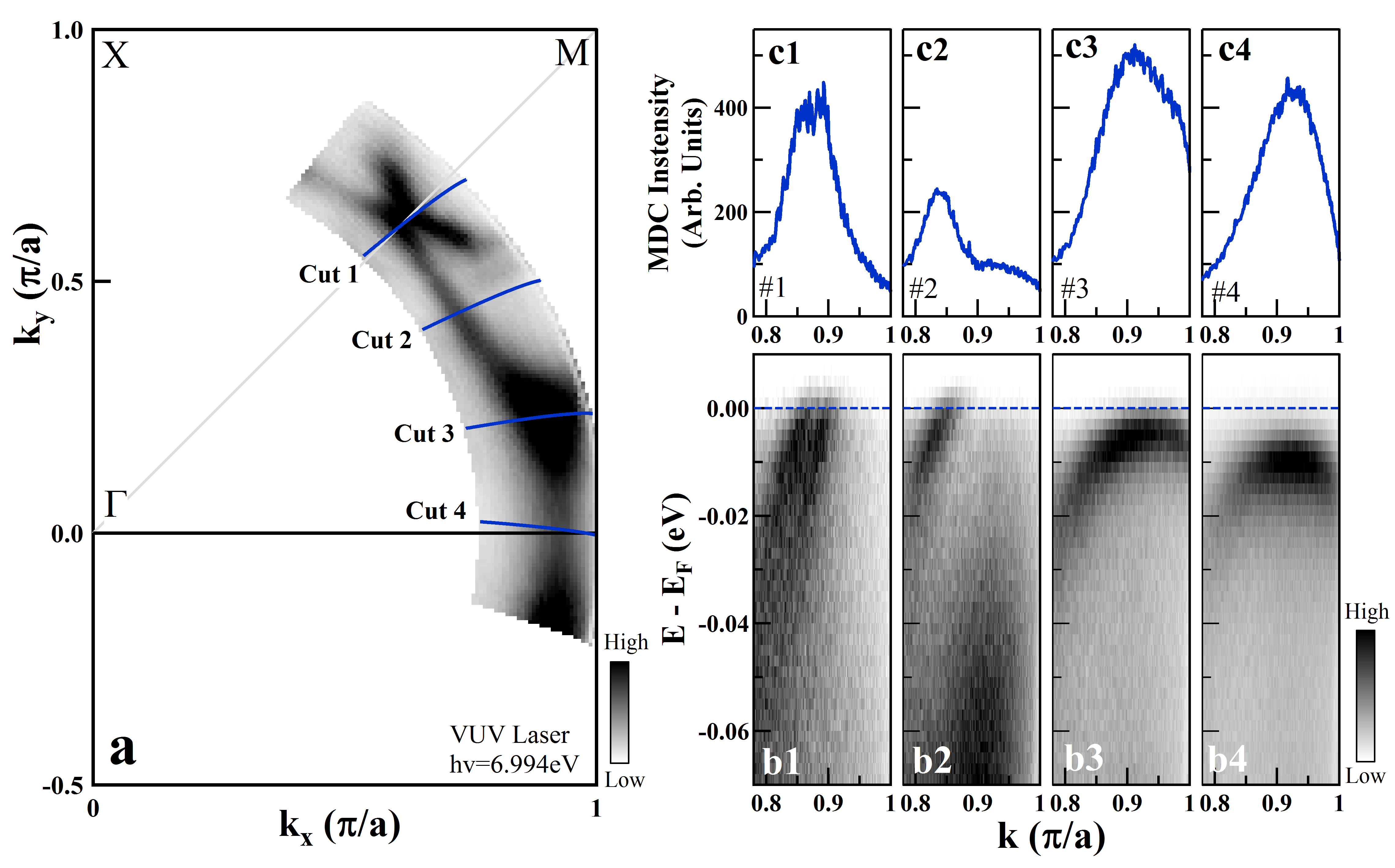}
\end{center}
\caption{Fermi surface and band structure of Sr$_2$RuO$_4$ measured by using $h\nu$=6.994 eV laser light source. The sample was cleaved at 20 K and measured at 20 K.  (a). Fermi surface mapping;  (b1-b4) show band structure for the four representative momentum cuts shown in (a). (c1-c4) shows the corresponding MDCs at the Fermi level.}
\end{figure}

In summary, we have systematically investigated various approaches to enhance the bulk components of the electronic structure in Sr$_2$RuO$_4$. We find that the cleaving temperature is not a key factor in suppressing the surface bands, as claimed before. Sample aging can help enhance bulk bands. VUV laser ARPES is an effective way to enhance the bulk bands due to the increase of photoelectron escape depth. These information are important in manipulating and detecting the surface properties of Sr$_2$RuO$_4$. In particular, the enhancement of bulk sensitivity, together with its super-high instrumental resolution of VUV laser ARPES, will be advantageous in investigating fine electronic structure and superconducting properties of Sr$_2$ruO$_4$ in the future.

\begin{flushleft}
\textbf{Acknowledgments}\\
We thank financial support from the NSFC (10734120 and 91021006) and the MOST of China (973 program No: 2011CB921703 and 2011CB605903).
\end{flushleft}

$^{*}$Corresponding author: XJZhou@aphy.iphy.ac.cn

\end{document}